\documentclass{technojournal}

\journalname{}

\title[On the origin of exponential growth]{On the origin of exponential growth in induced earthquakes in Groningen}

\correspinfo[]{mvp@sejong.ac.kr}

\author[a1]{Maurice H.P.M. van Putten}{\thanks{E-mail: \email{mvp@sejong.ac.kr}}}
\author[a2]{Anton F.P. van Putten}{\thanks{E-mail: \email{anton37.putten@wxs.nl}}}
\author[a3]{Michael J.A.M. van Putten}{\thanks{E-mail: \email{m.j.a.m.vanputten@utwente.nl}}}
\shortauthor{M. van Putten, A. van Putten and M. van Putten} 

\affiliation[a1]{Physics and Astronomy, Sejong University, Seoul, South Korea}
\affiliation[a2]{AnMar Research Laboratories B.V., Eindhoven, The Netherlands}
\affiliation[a3]{Clinical Neurophysiology, Medisch Spectrum Twente and University of Twente, Enschede, The Netherlands}

\usepackage{graphicx}
\usepackage{tabu,hhline,multirow}


\begin{document}

\maketitle
\begin{abstract}
The Groningen gas field shows exponential growth in earthquakes event counts around a magnitude M1 with a doubling time of 6-9 years since 2001. This behavior is identified with dimensionless curvature in land subsidence, which has been evolving at a constant rate over the last few decades {essentially uncorrelated to gas production.} We demonstrate our mechanism by a tabletop crack formation experiment. The observed skewed distribution of event magnitudes is matched by that of maxima of event clusters with a normal distribution. It predicts about one event $<$\,M5 per day in 2025, pointing to increasing stress to human living conditions.\\
\end{abstract}

\section{Introduction}
Induced earthquakes are a modern phenomenon associated with exploitation of natural energy resources or injection of waste 
\citep{ral76,nic90,fro12,eva12} at generally moderate strengths \citep[e.g][]{ell13}.
Recent studies in particular focus on the potential for a correlation between induced events and
fracking and waste water injection \citep{ker14,wei15}. Fig. 1 shows induced events in Groningen, The Netherlands, due to the exploitation of natural gas since 1959 \citep{kam14}.

Though moderate in magnitude, frequent events act as physical stressors to living conditions. As a relatively modern phenomenon, induced events pose some novel challenges for hazard analysis, 
beyond a mere extrapolation of the geophysical impact of natural earthquakes \citep{hou13}. In Groningen, half of the inhabitants are now living with a feeling of anxiety or fear and wish to move out, in part, by the stochastic nature of earthquakes \citep{fon09,gri04}. It has led to ameliorating policies (e.g. \cite{vol14}), generally seeking reduction in earthquake activity by limiting gas production, whose short term impact is uncertain \citep{nic90,bak05,nam13}.

Groningen data on reservoir pressure, land subsidence and yearly production of natural gas cover an area of 900 km$^2$ \citep{nam13}. Since 2001, earthquake event counts show exponential growth at seemingly `clock-work' precision (Fig. 1), {based on a logarithmic plot of event counts with remarkably good fit by a linear curve covering the year 2000 to the present. This observation applies to event counts with a choice of magnitude cut-off over a broad range $\mbox{M}_c=0-3$.  Collectively, they represent an energy release} $\dot{E}\simeq$\,3GJ in over one hundred registered events per year with a doubling time of 6-9 years (see also \cite{dos13}). The Groningen field apparently shows a robust principle at work with unknown but possibly long-term implications by exponential growth in counts.

Fig. 2 shows NAM production levels in GNm$^3$ per year and land subsidence, associated with a decline in reservoir {pressure} to below 100 bar at present, down from about 150 bar in 2000 and 350 bar in 1959. The statistical correlation {by Pearson coefficient} between event counts in Fig. 1 to production is nill, {indicative of no correlation (Fig. 2, lower left panel).}

These earthquakes arise from stresses induced  by a decline in well pressure \citep{hub59},  possibly at some distance away and delayed in time. Paradox Valley in Colorado \citep{ell13}  shows events at distances over ten kilometers with delays of a decade. The Salton Sea Geothermal Field shows, subsequent to initial exploitation, a correlation of seismicity with net extracted volume $\dot{V}$ (injected minus extracted volume) of about $10^6$ m$^3$ per month at 1-2.5 km depth \citep{bro13}. Large area induced seismicity points to large scale deformations that may be measured in land subsidence. 

\begin{figure}[h]
	\centerline{\includegraphics[scale=0.38]{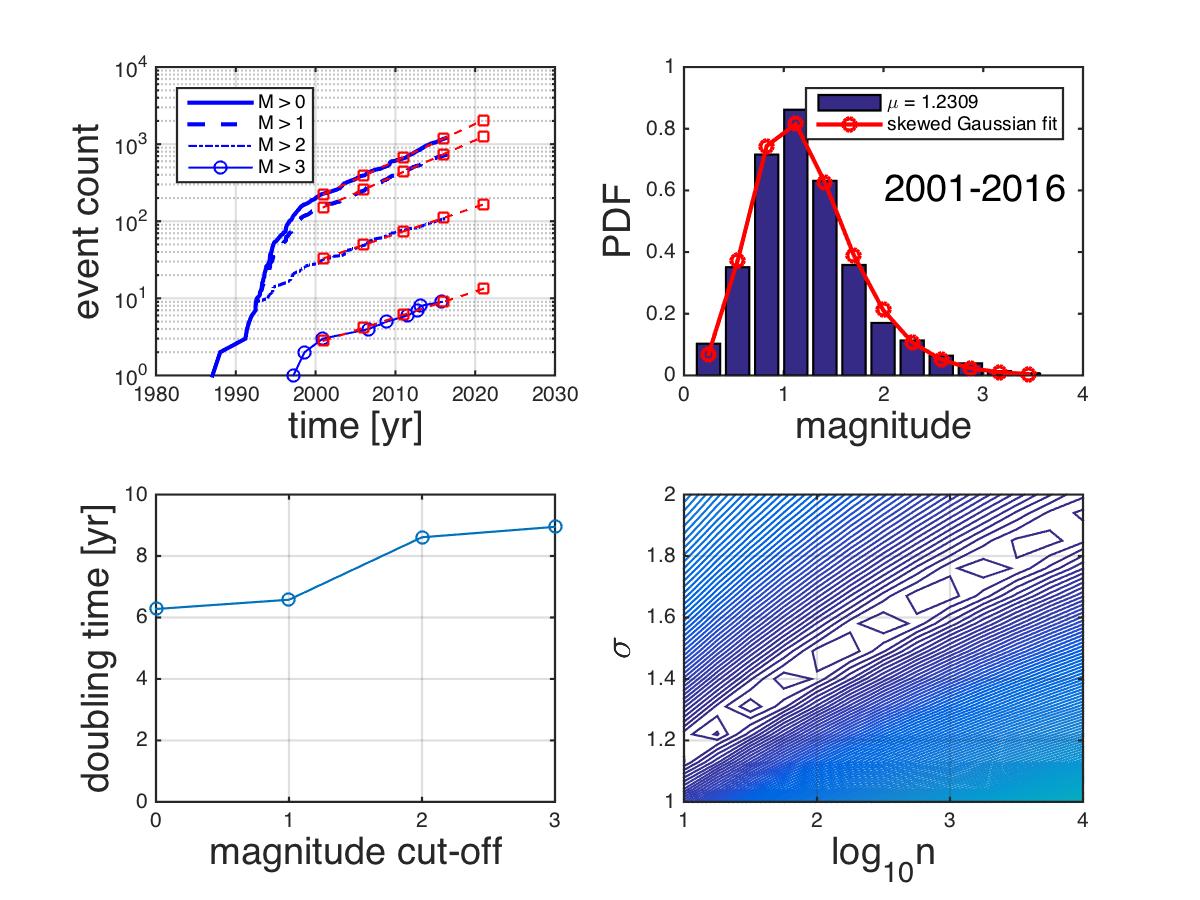}}
	\caption{(Left panels.) Exponential growth in event counts in Groningen that extrapolate to one M\,$>0$ event per day in 2025. (Right panels.) The distribution of magnitudes is skewed, here modeled by that of maxima of event clusters of size $n$, drawn from a normal distribution with dispersion $\sigma$. A contour plot of residual {\bf least square ($L_2$)} errors (bottom right) shows best-fit parameters $n\simeq 2250$, $\sigma\simeq 1.7$ (white strip).}
	\label{fig:GC1}
\end{figure}
\begin{figure}[h]
	\centerline{\includegraphics[scale=0.38]{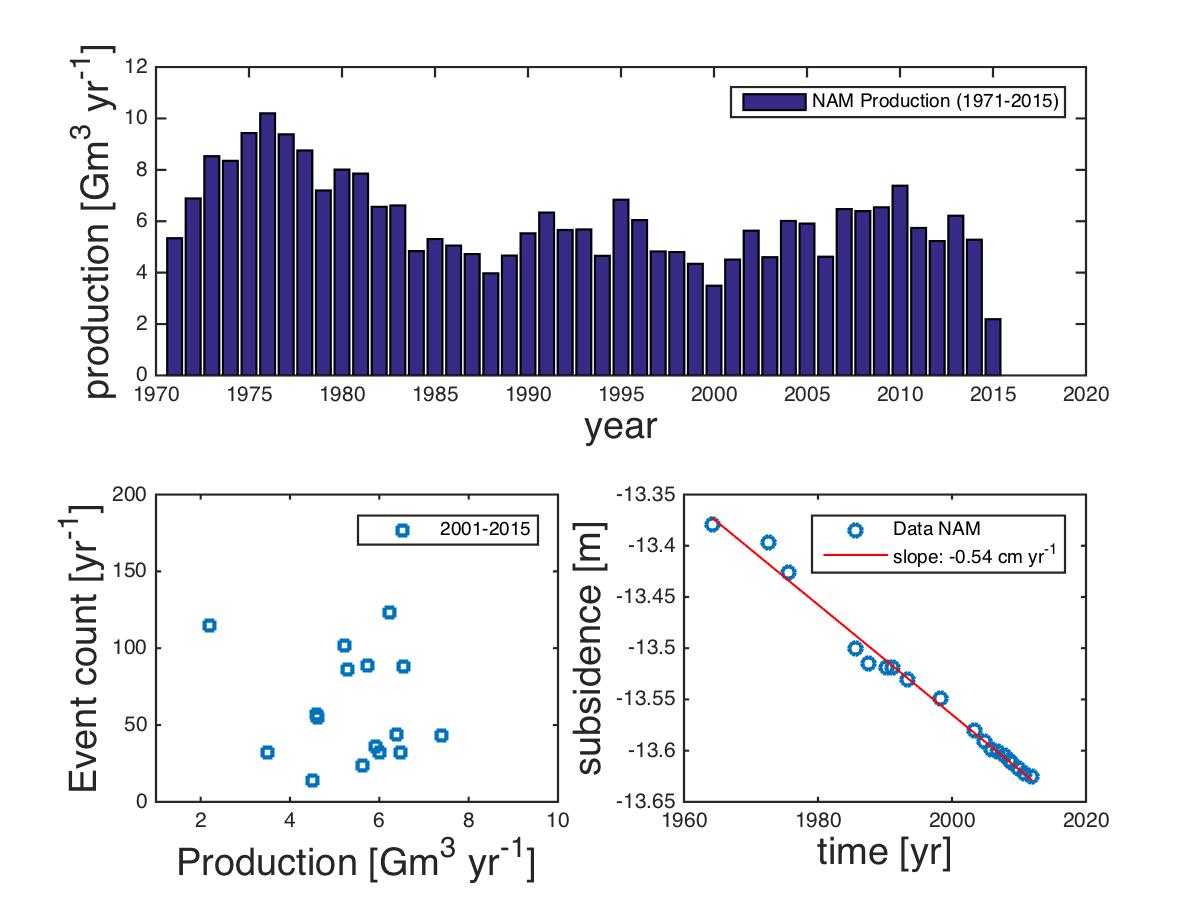}}
	\caption{(Top panel.) NAM production. (Left bottom.) Scatter plot of event counts (Fig. 1) versus production over 2001-2015. The absence of a correlation (Pearson coefficient 0.0816)  evidencies a lack of sensitivity to NAM production. (Right bottom.) Subsidence increases essentially linear in time since 1970. (Data from \cite{nam13}).} 
	\label{fig:GC2}
\end{figure}

Here, we propose a novel physical mechanism for exponential growth in event counts in induced earthquakes in response to dimensionless curvature of land subsidence, demonstrated by a table top experiment. Based on dimensionless variables, it is insensitive to details of geophysical composition. The skewed distribution in magnitudes is modeled by maxima of event clusters, {that will be used for a statistical forecast on event count and maximal magnitude in the coming decade.}

\section{Exponential growth from curvature induced strain}

Groningen land subsidence is $\dot{h}\simeq 0.5$ cm yr$^{-1}$ (Fig. 2, right lower panel) with a delay in response to a reduction in volume $V$ due to a gradual compactification in the reservoir associated with a steady decrease in reservoir pressure \citep{nam13}. Compactification $\dot{V}\simeq 10^6$ m$^{-3}$ per month is roughly similar to that in aforementioned Salton Sea Geothermal Field. It represents a mechanical energy input of $\dot{E}_0\simeq$\,0.3TJ per year, whereby $\dot{E}/\dot{E}_0\simeq 10^{-5}$.

Subsidence gives rise to curvature-induced strain, inevitably leading to the build up of internal stresses. At critical values, materials fail by cracks or slip, described by the Mohr-Coulomb law \citep{moh76,hub59}. Subsidence in Groningen is taken the shape of a shallow bowl of radius $r\simeq$ 15 km (\cite{nam13}, Fig. 3). Relative to its radius of curvature $R$, the dimensionless curvature $\rho=r/R\simeq 4\times 10^{-5}$ should be representative for large scale internal strain. For small perturbations in the absence of cracks and slides, internal stresses increase with strain induced at a well, satisfying a linear elliptical equation subject to free surface conditions on the land above and maximal strain and pressure gradients in the well at the reservoir depth of about 3 km. Following Fig. 3, internal strain hereby scales linearly with subsidence $\delta h = R(1-\cos\theta)\simeq R\theta^2/2$, $R\sin\theta =r$, whereby $\delta h \simeq r^2/2R$, i.e., 
\begin{eqnarray}
\frac{\delta h}{r}\simeq \frac{1}{2}\rho.
\label{EQN_hrho}
\end{eqnarray} 
Glitches appear by cracks and slides when internal strains $\epsilon$ reach a critical value $\epsilon_c$ with associated critical values for (\ref{EQN_hrho}). Such impulsive release of internal stress is followed by relaxation to sub-critical values of $\epsilon$.

\begin{figure}[h]
	\centerline{\includegraphics[scale=0.40]{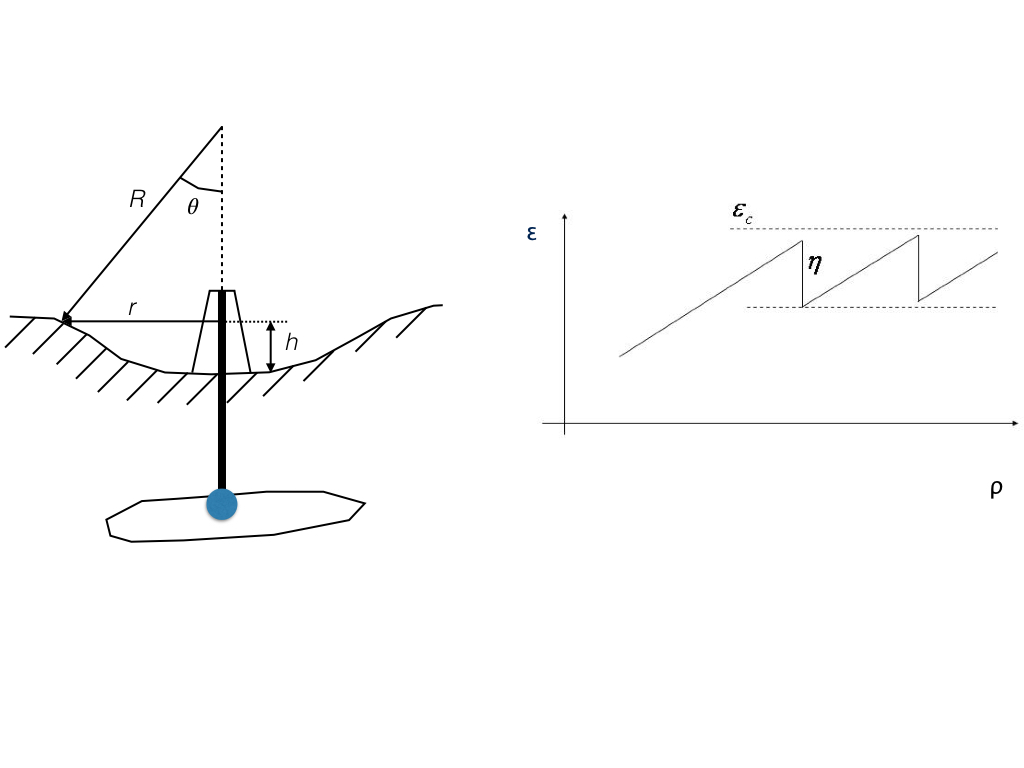}}
	\caption{(Left.) Geometry of land subsisdence around a well. Internal strain $\epsilon$ scales linearly with dimensionless curvature $\rho=r/R$ due to land subsidence. (Right.) Evolution of $\epsilon$ as a function of $\rho$ to critical values $\epsilon_c$ with glitches and relaxation by $\eta$ to sub-critical values.}
	\label{fig:S0}
	\label{fig:T0}
\end{figure}

\begin{figure}[h]
	\centerline{\includegraphics[scale=0.3]{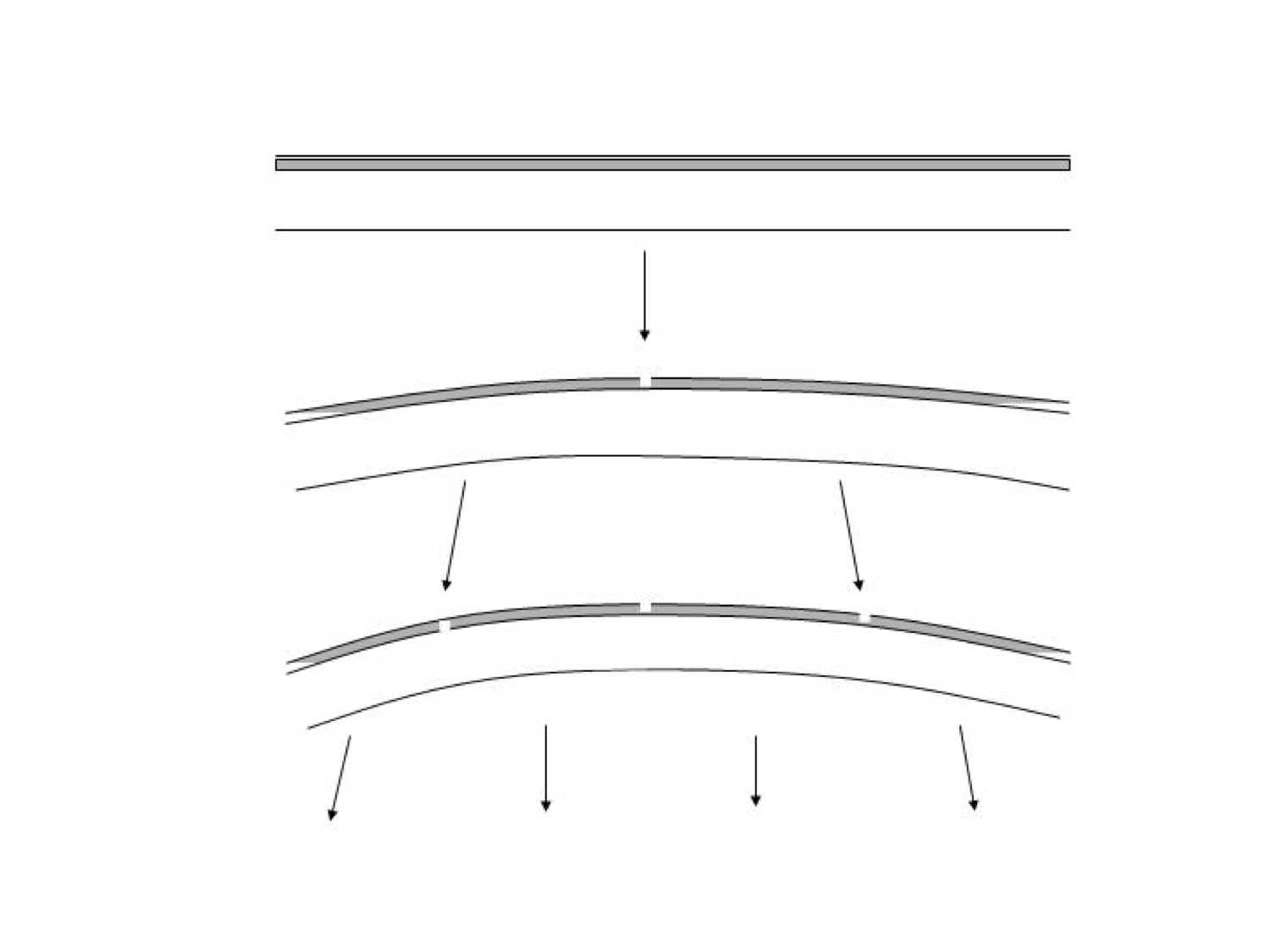}}
	\caption{Schematic of a tabletop experiment on cracks, realized in aluminum foil glued to a flexible substrate of length $L\simeq 20$\,cm subject to bending. The first crack induced by dimensionless curvature $L/R$ defined by the radius of curvature $R$ forms at a critical strain $\epsilon=\epsilon_c$. Each such partition produces two sections, followed by relaxation to a stressed but stable state at a strain $\epsilon_c-\eta$. Further bending rebuilds stress back to $\epsilon_c$, causing both sections to crack, again followed by relaxation.}
	\label{figE_1}
\end{figure}
\begin{figure}[h]
	\centerline{\includegraphics[scale=0.3]{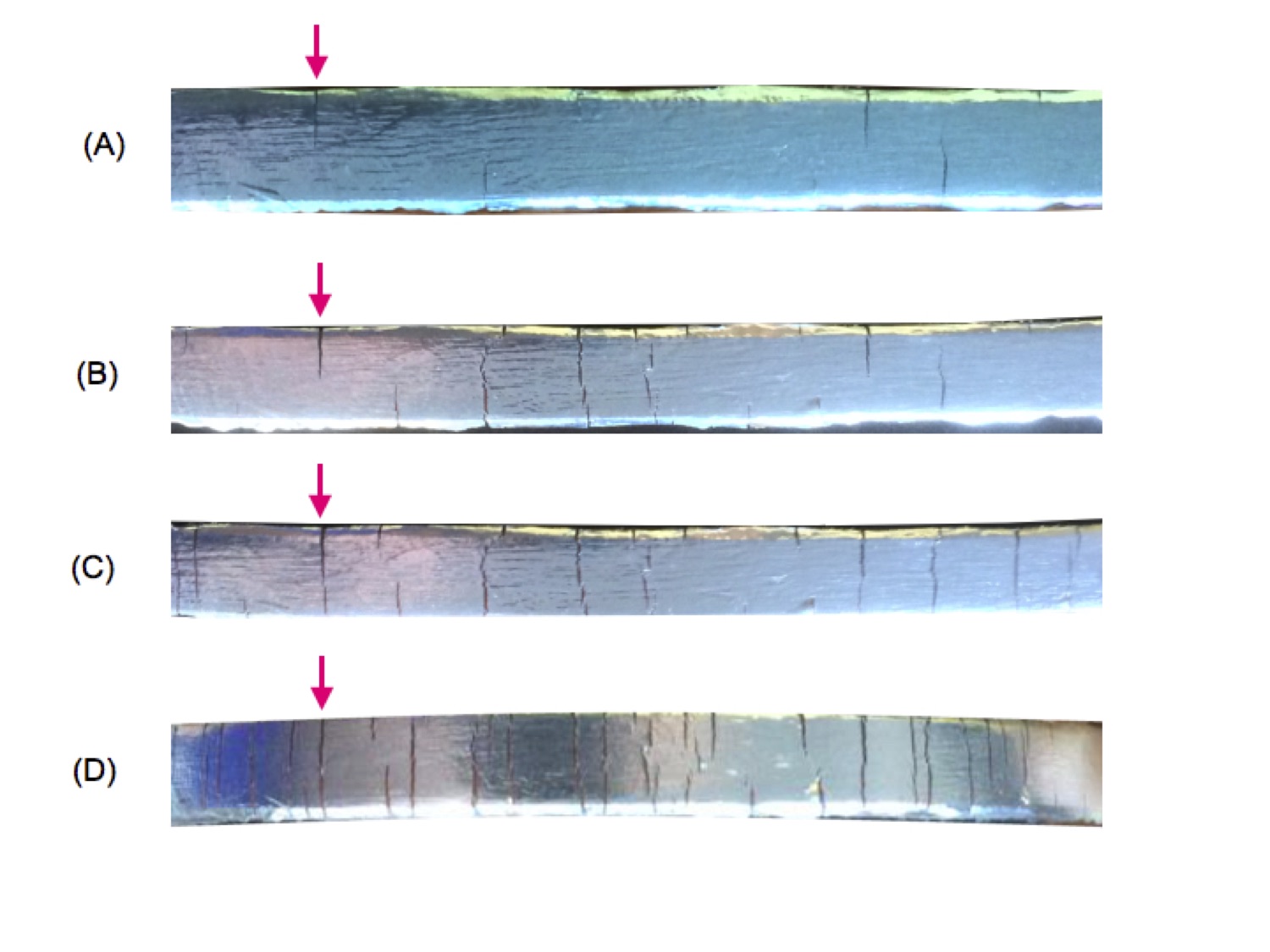}}
	\caption{Snapshots of crack formation in the aluminum foil subject to bending, increasing from A-D. Arrows point to a single crack for reference.}
	\label{figE_2}
\end{figure}
\begin{figure}[h]
	\centerline{\includegraphics[scale=0.3]{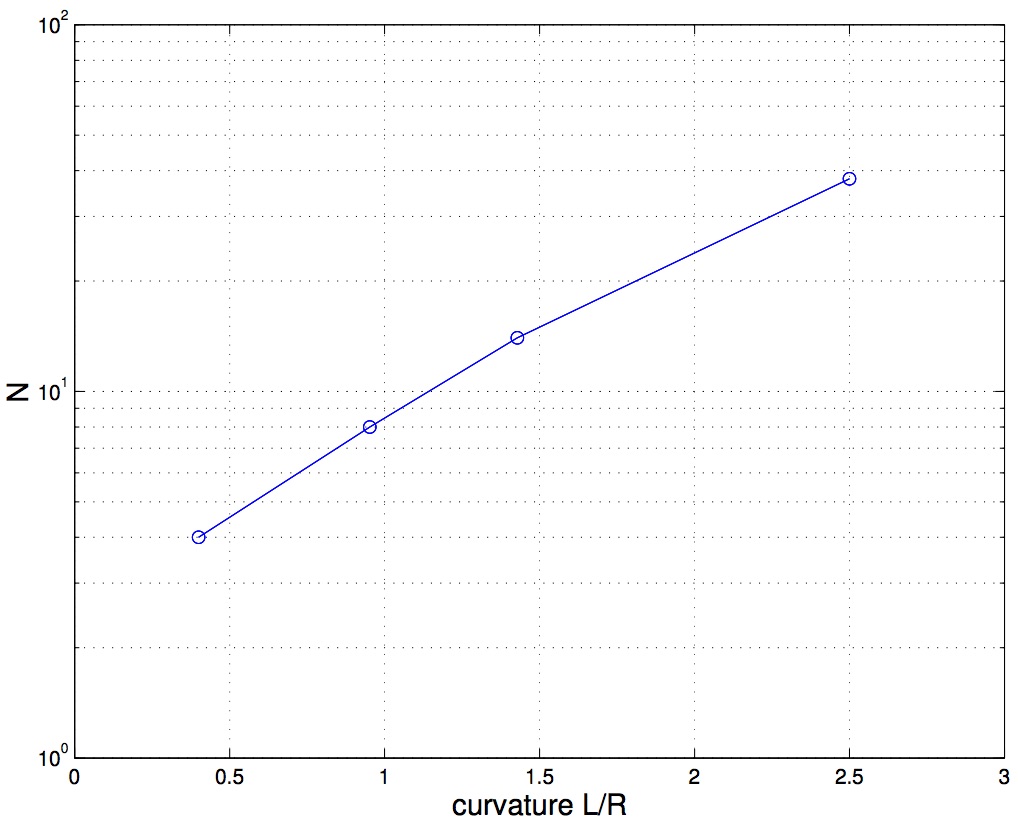}}
	\caption{The observed number of cracks $N$ shows an approximately exponential growth with $\rho$.}
	\label{figE_3}
\end{figure}

Glitches partition a region into two followed by a relaxation $\eta$ to a sub-critical state, satisfying 
\begin{eqnarray}
0 < \eta < \epsilon_c.
\end{eqnarray}
Continuing subsidence $\delta h/r < 0$ increases $\eta$, back to $\epsilon_c$ over time.
In approaching $\epsilon_c$, regions in the new partition crack or slice into two once again. A time-evolving strain $\epsilon=\epsilon(t)$ is hereby restricted to the range (Fig. 3)
\begin{eqnarray}
\epsilon_c-\eta < \epsilon < \epsilon_c.
\label{EQN_eps}
\end{eqnarray}
As $\rho$ continues to grow by subsidence, a shallow bowl forms in the region around the natural gas reservoir. Induced strain $\epsilon$ hereby extends approximately uniformly over the entire area of the bowl and (\ref{EQN_eps}) produce exponential growth in event count
\begin{eqnarray}
N = N_0 e ^{\alpha \rho},
\label{EQN_N}
\end{eqnarray}
where $\alpha$ is a dimensionless constant. For Groningen, $N\simeq 10^3$ today (since 2001), $\dot{N}/N\simeq 0.1$, $\rho\simeq 4\times 10^{-5}$ and $\dot{\rho}\simeq 6.7\times 10^{-6}$, give the estimates $\alpha\rho\simeq 6$ and $N_0$ is of order a few, as expected. 

Figs. 4-6 shows a table top experiment demonstrating (\ref{EQN_N}). This $\epsilon$ excess of $\epsilon_c$ is visible in cracks in aluminum foil, here glued to a flexible substrate as a function of dimensionless curvature $\rho$. In this experiment, $\eta$ is essentially equal to the critical value $\epsilon_c$.  It serves to illustrate that, in general, $\eta$ need not be small and can be on the order of $\epsilon_c$. 

Our mechanism (\ref{EQN_N}) is completely dimensionless. It does not explicitly depend on the details of internal length, area or volume or geophysical structure. Independent of dimension, it has implications also to three dimensional microfracturing experiments \citep{sch68} on induced fractures. This applies especially, since the latter are typically limited to a few hundred meters \citep{dav12,lac13}, which is rather local relative to the scale depth of about 3 km of the Groningen reservoir. Given that subsidence in the Groningen field increases at an approximately constant rate in time (Fig. 2), we therefore identify the exponential trend (Fig. 1) with (\ref{EQN_N}).

Our model does not address time delay between exploitation and evolution of subsidence. Time delays are notoriously varied in different locations, and their origin remains somewhat mysterious especially so when delays are long. For Groningen, the lack of any correlation between gas exploitation and observed event counts (Fig. 2) over the course of one decade is remarkable. It points to long-lived delayed events, likely on the order of a decade.

\section{Parameter-free statistical forecast of magnitude limit}

For Groningen, we attribute induced events to sudden release of energy in clusters of essentially contemporaneous events, derived from normal distributions in $\eta$ and $\epsilon_c$. The magnitude of an earthquake (as reported) is then the maximum of the $\eta$ in such cluster (cf. the composite source model of \citep{fra91}). 

In general, a collection of $n$ independent samples of a normal variable $x$ with mean zero and dispersion $\sigma$ produces a truncated Gaussian distribution, {whenever $n$ is finite, and maxima thereof satisfy a skewed distribution. To see this, the probability density function (PDF) of the absolute values satisfies \citep[e.g.][]{van14,van16})}
\begin{eqnarray}
p(x) = \frac{n}{\sigma}\sqrt{\frac{2}{\pi}}\,\mbox{erf}(y)^{n-1} e^{-y^2}
\label{EQN_SG}
\end{eqnarray}
with $y=x/{\sqrt{2}\sigma}$, derived from the probability $P(<x)=\int_0^x p(s)ds$ that all $n$ samples are less than $|x|$. 
Fig. 1 shows that (\ref{EQN_SG}) accurately matches in the observed distribution of Groningen for $n=2250$ with $\sigma=1.7$, which is an almost degenerate minimum along a line in the $(n,\sigma)$ plane. Following Fig. 3, the observed magnitudes are the maxima in $\eta$  associated with the random variable $x=\epsilon_c$. Our match to the distribution of the Groningen events involves a shift downwards from the mean $\mu=6.4$ in $x$.

Figs. 7-8 shows that our matching PDF derived from (\ref{EQN_N}) has a decay that is slightly steeper than exponential with associated slowly varying Gutenberg-Richter $b$ value 1.1-1.3, consistent with earlier estimates that depend on a choice of magnitude cut-off $M_c$ \citep{ish39,gut44,gut54,sch05}. For statistical forecasting, our PDF takes into account weak deviations from exponential decay and obviates the need for using a choice of magnitude cut-off.

\section{Summary and outlook}

\begin{figure}[h]
	\centerline{\includegraphics[width=150mm,height=100mm]{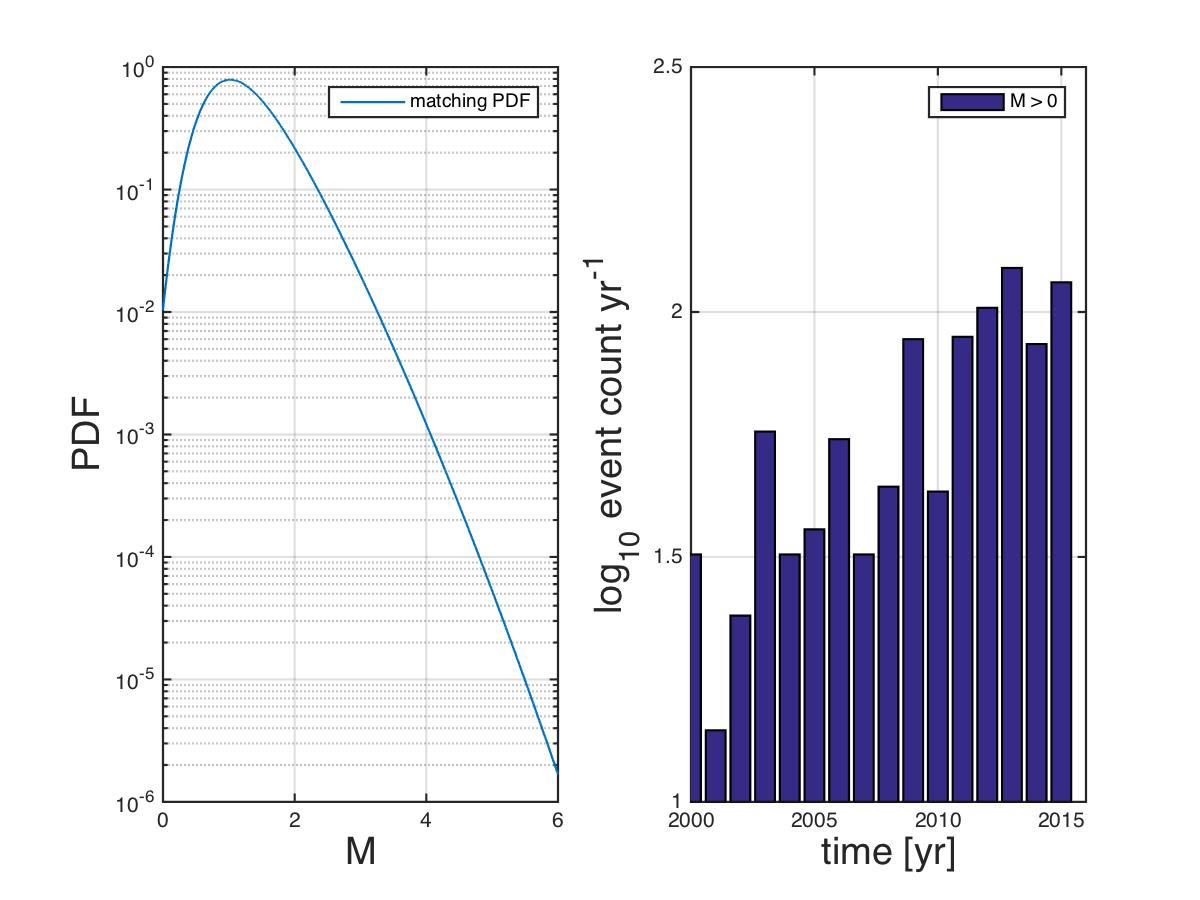}}
	\caption{Overview of a statistical forecast based on the optimal fit of the PDF in (\ref{EQN_SG}) to the observed skewed distribtion of events in Fig. 1.}
	\label{figF_1}
\end{figure}
\begin{figure}[h]
	\centerline{\includegraphics[scale=0.25]{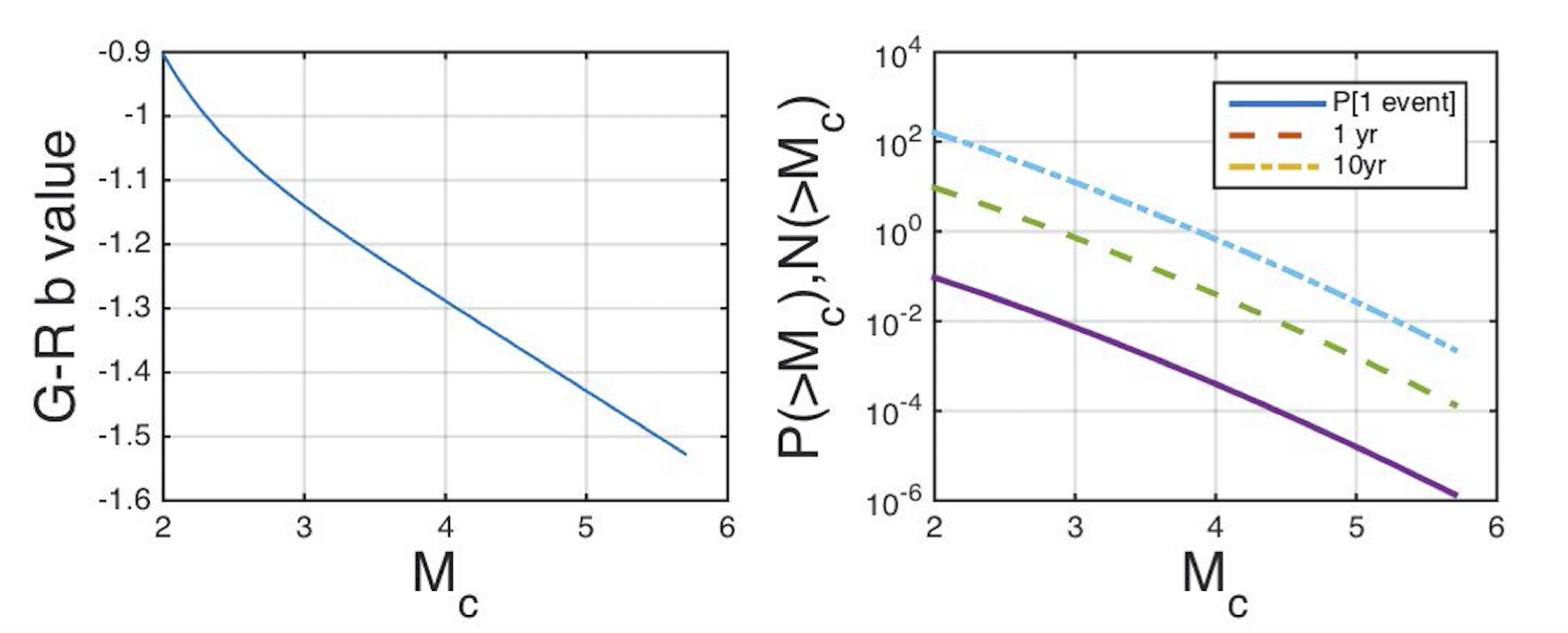}}
	\caption{The PDF in (\ref{EQN_SG}) to the observed skewed distribution of events in Fig. 1 has an associated Gutenberg-Richter $b$-value $-1.1 < b < -1.3$ $(3<M_c<4$) (left).
		Extrapolation of the current yearly event rates (right top) and the cumulative probability $P(>M_c)$ derived from the fit (\ref{EQN_SG}) shows $N(>M4)\simeq 1$ and $P(>M5)\simeq 1\%$ over the next decade (right).}
	\label{figF_2}
\end{figure}

Exponential growth in Groningen event counts is a remarkable man-made phenomenon. We attribute it to curvature-induced strain in light of the relatively constant rate of land subsidence observed over the last few decades. Based on dimensionless variables and demonstrated by experiment, our mechanism is robust and does not depend on the detailed geophysical struture of the Groningen soil composition.

{Fig. 1 shows a doubling time of 6-9 years in total event count and the observed PDF of event magnitudes with positive skewness. Figs. 7-8 show a matching PDF of maxima of event clusters (\ref{EQN_SG}). This matching PDF shows a Gutenberg-Richter value that, in absolute value, increases steeply with the cut-off $M_c$ (Fig. 8). While it assumes typical values $|b|\simeq1.15-1.3$ for $M_c\simeq 3-4$, $|b|$ increases to about 1.5 for $M_c\simeq 5-6$. As a result, we arrive at the following forecast for 2025: 
\begin{itemize}
\item An event rate N($>$M0) of one per day;
\item N($>$M4)$\,\,\simeq 1$ per year;
\item A probability of 1\% for one $>$M5 event per year. 
\end{itemize}
The relatively small probability of the latter is due to aforementioned increase in $|b|$-value in the tail of the PDF, larger than typical estimates based on moderate cut-off magnitudes $M_c$.}

Based on $\dot{E}/\dot{E}_0\simeq 10^{-5}$, there is no energetic limit to the observed exponential growth for decades to come. (Theoretically, it can continue for about one century.) An apparent slow large scale creep delayed relative to production is consistent with a lack of a correlation of yearly subsidence with fluctuations in gas production and reservoir pressure, whose physical origin, however, remains somewhat mysterious. It suggests a significant time delay time on the order of a decade or more, that challenge the formulation of effective policies and strategies to ameliorate this bleak outlook as a well-populated area (cf. \cite{eck06,pet08}).

While the number of earthquakes is rapidly increasing with the potential for significant continuing stresses to human living conditions, their strength probably remains below M5. 
For Groningen, we propose that the exponential growth in event counts is included in hazard analysis as a general stressor to living conditions. Also, the relatively high frequency spectrum of induced motions is potentially relevant, as it falls in the range of eigenfrequencies of around 0.1-25 Hz, {where the higher frequencies are expected to be relevant to homes in the Groningen area.}
 (High frequency earthquakes may be found, for instance, in the spectrum of induced events at the Hellisheidi Geothermal Powerplang in Iceland \citep{hal12} peaks at 10 Hz.) These relatively high frequencies $f$ are commonly attributed to the scaling $f\propto M^{-1/3}$ with magnitude \citep{aki67,bun70,kam75,and03,aki80}. 

\section{Acknowledgments.} The authors thank the anonymous reviewer for constructive comments on the manuscript, and the first author thanks Tae Woong Chung for stimulating discussions. Data in Figs. 1-2 are from KMNI (http://rdsa.knmi.nl/dataportal) and the Nederlandse Aardolie Maatschappij (NAM, http://www.namplatform.nl/feiten-en-cijfers/gaswinning, last accessed March 1 2016). This research was partially supported by a Sejong University Faculty Research Fund and by the Korean National Research Foundation under Grants 2015R1D1A1A01059793 and 2016R1A5A1013277.

\clearpage


\end{document}